\documentclass[12pt,preprint]{aastex}
\usepackage{apjfonts}
\slugcomment{{\sc Accepted for publication in ApJ Letters:} August, 2009}

\usepackage{epsfig}
\usepackage{lscape,graphicx}
\usepackage{url}

\shorttitle{Kinematics and Metallicities in BooIII}
\shortauthors{Carlin et al.}

\begin{document}

\title{Kinematics and Metallicities in the Bo\"{o}tes III Stellar Overdensity:  a Disrupted Dwarf Galaxy? \footnote{Observations reported here were obtained at the MMT Observatory, a joint facility of the Smithsonian Institution and the University of Arizona.}}

\author{
Jeffrey L. Carlin\altaffilmark{1},
Carl J. Grillmair\altaffilmark{2},
Ricardo R. Mu\~noz\altaffilmark{3},
David L. Nidever\altaffilmark{1}, \&
Steven R. Majewski\altaffilmark{1}}

\altaffiltext{1}{Department of Astronomy, University of Virginia, P. O. Box
400325, Charlottesville, VA 22904-4325, USA (jcarlin, dln5q, srm4n@virginia.edu)}

\altaffiltext{2}{Spitzer Science Center, 1200 East California Boulevard,
Pasadena, CA 91125, USA (carl@ipac.caltech.edu)}

\altaffiltext{3}{Yale University, Astronomy Department, P.O. Box
208101, New Haven, CT 06520-8101, USA (ricardo.munoz@yale.edu)}

\begin{abstract}

We report the results of a spectroscopic study of the Bo\"{o}tes III
(BooIII) stellar overdensity carried out with the Hectospec multifiber
spectrograph on the MMT telescope. Radial velocities have been
measured for 193 BooIII candidate stars selected to have magnitudes
and colors consistent with its upper main sequence and lower red giant
branch, as well as a number of horizontal branch candidates.  From 20
identified candidate BooIII members, we measure a systemic velocity of
$V_{\sun}=197.5\pm3.8$ km s$^{-1}$ and a velocity dispersion of
$\sigma_{\rm o}=14.0\pm3.2$ km s$^{-1}$.  We use the somewhat large
velocity dispersion and the implied highly radial orbit, along with
morphological evidence from \citet{Grillmair2009} and stellar
abundances, to argue that BooIII is likely the first known object
observed in a transitional state between being a bound dwarf galaxy
and a completely unbound tidal stream.

\end{abstract}


\section{Introduction}

Over the last few years and thanks to the advent of large-area
photometric surveys such as the Two Micron All Sky Survey (2MASS) and
the Sloan Digital Sky Survey (SDSS; \citealt{SDSS_DR7_2009}), the
substructured nature of the Milky Way halo has been clearly revealed.
In recent years the number of known Galactic dwarf spheroidals (dSphs)
has more than doubled (e.g., \citealt{Willman2005};
\citealt{Belokurov2006_Bootes,Belokurov2007_Quintet};
\citealt{Zucker2006_UMaII,Zucker2006_CVn};
\citealt{Grillmair2006_Orphan}), and numerous stellar streams have
been identified (e.g. \citealt{Ibata2001}; \citealt{Majewski2003};
\citealt{RochaPinto2004}; \citealt{Yanny2003};
\citealt{Belokurov2006_FOS};
\citealt{Grillmair2006_Orphan,Grillmair2006_ACS,Grillmair2009};
\citealt{GD2006}).  These discoveries lend significant support to
currently favored cold dark matter cosmological models, which predict
that spiral galaxies form hierarchically from the merger and accretion
of smaller subsystems (e.g.,
\citealt{Bullock_etal2001},\citealt{Bullock_Johnston2005}).

One exciting aspect of these discoveries is that they have
revealed a previously unknown population of ``ultra-faint" dwarf
galaxies which are claimed to contain the highest dark matter
fractions reported to date, with mass-to-light ratios above $1,000$
(M/L)$_{\sun}$ in some cases (e.g., \citealt{Munoz2006};
\citealt{Simon_Geha2007}; \citealt{Strigari2008}), but luminosities as
low as $L_{V}\sim300$ L$_{\sun}$ \citep{Martin2007}.  A few of these
satellites, Coma, Segue 1, Segue 2 and Bo\"otes II, have properties
closer to those of globular clusters than those of ``classical" dSphs
(e.g., half-light radii between 30 and 70 pc), and have been suggested to
be satellites of satellites \citep{Belokurov_etal2009}.

In this Letter we present the results of a spectroscopic survey of
another potential dwarf satellite,
the Bo\"{o}tes III stellar overdensity (BooIII; \citealt[][hereafter
{G09}]{Grillmair2009}).  This peculiar object, located at a distance
of 46 kpc and with $V$-band absolute magnitude of
$M_{\rm V}=-5.8\pm0.5$ \citep{Correnti2009}, has a roughly power-law
surface brightness profile and an irregular shape.  Our study suggests
that BooIII is likely a dwarf satellite in a transitional state
between a bound entity and an unbound stellar stream. In this context,
BooIII resembles the objects described by \citet{Kroupa1997} but
likely with a high dark matter content.

\section{Spectroscopic Observations and Data Reduction}

Radial velocities (RVs) were measured from spectra obtained in
2009 February with the Hectospec multifiber spectrograph
\citep{Fabricant2005} mounted at the f/5 focus on the 6.5 m MMT.
Targets were selected from the SDSS stellar database to lie on or near
the BooIII color-magnitude sequence from \citeauthor*{Grillmair2009},
in a magnitude range of 18.5$ < g < $22.5.  We also included candidate
blue horizontal-branch (BHB) stars that are near the BHB locus in
Figure 13 of \citeauthor*{Grillmair2009}, bringing the total number of
assigned targets in the $\sim$1 degree Hectospec field of view to 227.
Figure 1 shows the $g - i$ versus $g$ color-magnitude diagram (CMD) for
all stars within 0.5$^{\circ}$ of the BooIII center, as well as an old,
metal-poor isochrone chosen to be similar to that of M15 (the cluster
which guided our target selection).

We used the 600 gpm grating, centered at 5800\AA, to give a working
wavelength range of 4550-7050\AA~at a dispersion of $\sim$0.55
\AA$~$pixel$^{-1}$ (2.1 \AA~resolution).  This region was selected to
include the H$\beta$, Mg triplet, Na D, and H$\alpha$ spectral
features.  Queue-observed integrations totaled 3 $\times$ 1800 s on UT
2009 February 3, and 6 $\times$ 1800 s on UT 2009 February 20, for a
summed exposure of 4.5 hr on this field.  Sky fibers (57 in all)
were assigned throughout the field, distributed so that a number of
them would fall within each of the two CCD chips of the Hectospec
system.  A total of 193 target spectra had star-like spectra (there
were a number of background QSOs/AGN), with sufficient signal-to-noise ratio (S/N $>$ 10)
for RV measurement.

The Hectospec data were reduced using an external version of the SAO
``SPECROAD'' reduction pipeline \citep{Mink2007} written by Juan
Cabanela, called
ESPECROAD\footnote{\url{http://iparrizar.mnstate.edu/~juan/research/ESPECROAD/index.php}}.
The pipeline automates many reduction steps, including
bias-correction, flat-fielding, cosmic-ray rejection, fiber-to-fiber
throughput adjustments, and sky subtraction.  Wavelength calibration
was performed manually using sets of three combined HeNeAr calibration
lamp exposures from each night of observing.

We derived RVs using the IRAF task FXCOR to cross-correlate object
spectra against two template G8V RV standards, each observed three
times.  We first correlated the standard spectra against each other,
and found that to achieve agreement between our measurements and
published RVs\footnote{From the Geneva Radial Velocity Standard Stars
at \url{http://obswww.unige.ch/~udry/std/}} for these stars, it was
necessary to shift all RVs measured against one of them by 6 km
s$^{-1}$.  After doing so, the measured RVs for all six standard
spectra agree with published values to within 1.0 km s$^{-1}$.  To
minimize the effects of noise for spectra with lower S/N, the
cross-correlation was restricted to the regions around the H$\alpha$,
Mg triplet, and H$\beta$ lines.  The standard deviation of six RV
measurements for most objects (after applying the 6 km s$^{-1}$ shift
discussed above) was less than 1 km s$^{-1}$, increasing to $\sim$3 km
s$^{-1}$ at $g\sim$22.

To estimate RV uncertainties, we separated our nine individual object
exposures into three sets of three summed exposures, totaling 1.5
hr of exposure each.  We then measured RVs in each of these three
independent frames, and fit a curve to the run of standard deviation
of the three measurements as a function of magnitude.  We used this
fit to the trend of $\sigma_{\rm RV}$ versus $g$ for the brighter stars
in our sample to assign an RV uncertainty for each star we observed
based on its magnitude (extrapolating the curve for fainter stars
having insufficient signal to measure an RV from the 1.5 hr
exposures).  Velocity uncertainties vary from $\sigma_{\rm Vhelio}
\sim 3$ km s$^{-1}$ at $g = $18.5 to $\sim$10 km s$^{-1}$ at $g =$ 21
(and higher at fainter magnitudes).

\section{Spectroscopic Results}

\subsection{Bo\"{o}tes III Membership}

\begin{figure}
\epsscale{0.85}
\plotone{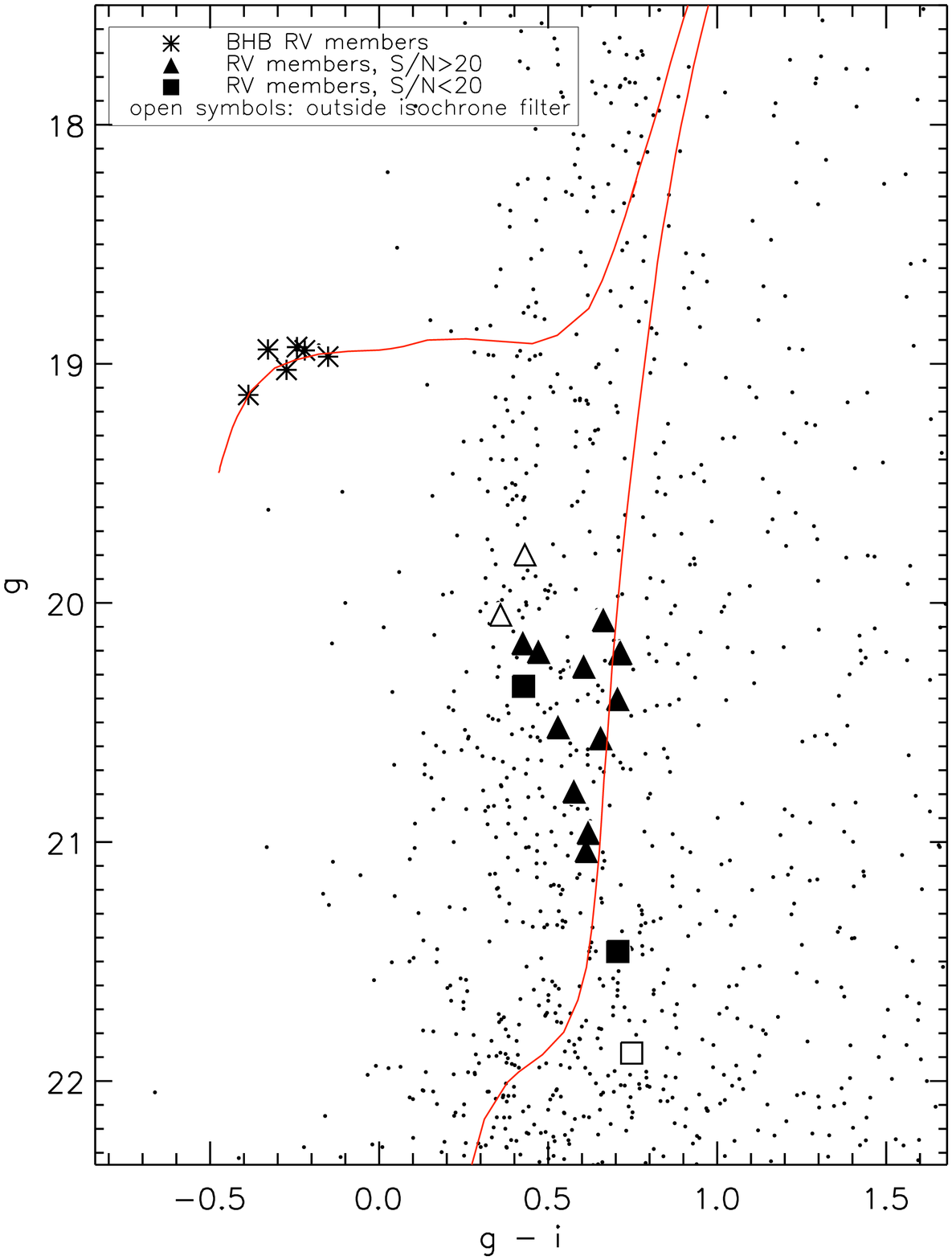}
\caption{SDSS color-magnitude diagram of a 30 arcmin radius about
the nominal BooIII center, drawn from the same data points as Figure 13
of \citet{Grillmair2009}.  The solid line is an old, metal-poor
([Fe/H] = -2.30) isochrone from \citet{Girardi2000}, shifted to a
distance of 46.0 kpc.  Large symbols represent all stars within the
3-$\sigma$ RV selection centered at $\sim$197.5 km s$^{-1}$; asterisks
represent the six BHB candidates that were in our spectroscopic
sample, triangles are RV members whose spectra had S/N $>$ 20, and
squares those having S/N $<$ 20.  Filled symbols and asterisks are
candidate RV members that are also within 0.25 magnitudes of the
metal-poor isochrone in the CMD; these constitute our final BooIII
sample.}
\end{figure}

Figure 2 shows RVs for all stars in our sample
with reliable measurements (S/N $>$ 10).  For comparison, we
show (red dot-dashed curve) a scaled RV distribution from $\sim$100
realizations of the Besan\c{c}on Galactic model \citep{Robin2003}
along this line of sight, covering the same magnitude and color
ranges as our targets.  Since this field is at high Galactic latitude
($b\sim75.4^{\circ}$), blue stars at faint magnitudes are mostly Milky
Way halo main sequence turnoff stars, with a large ($\sim$100 km
s$^{-1}$) velocity dispersion.  This halo distribution is well matched
by our data set. However, two significant excesses with respect to the
expected halo distribution, at V$_{\rm helio}$ of -180 km s$^{-1}$ and
$\sim$200 km s$^{-1}$, are visible in the histogram.  To determine
whether one of these RV signatures is associated with BooIII we check
the velocities of the six candidate BooIII BHB stars we have included
from the ``clump'' in Figure 13 of \citeauthor*{Grillmair2009} (asterisk
symbols in Figure 1).  Interestingly, all six of these stars have
velocities within the $\sim$200 km s$^{-1}$ peak, with a mean value of
V$_{\rm helio} = 195.6 \pm 7.2$ km s$^{-1}$ and dispersion 13.7 km
s$^{-1}$.  Since none of the BHB candidates have velocities within the
negative RV peak, we identify this positive, high-velocity peak with
the BooIII kinematical signature.

Having thus identified BooIII's systemic velocity, we select other
stars within 3-$\sigma$ of this BHB velocity peak as candidate BooIII
members.  These stars are highlighted in the $g - i$ versus $g$ CMD in
Figure 1.  The good agreement between the isochrone for an old,
metal-poor ([Fe/H] = -2.30) population and the BooIII RV candidates
suggests that we are indeed seeing the red giant branch of an old,
metal-poor, comoving population of stars.

\begin{figure}
\plotone{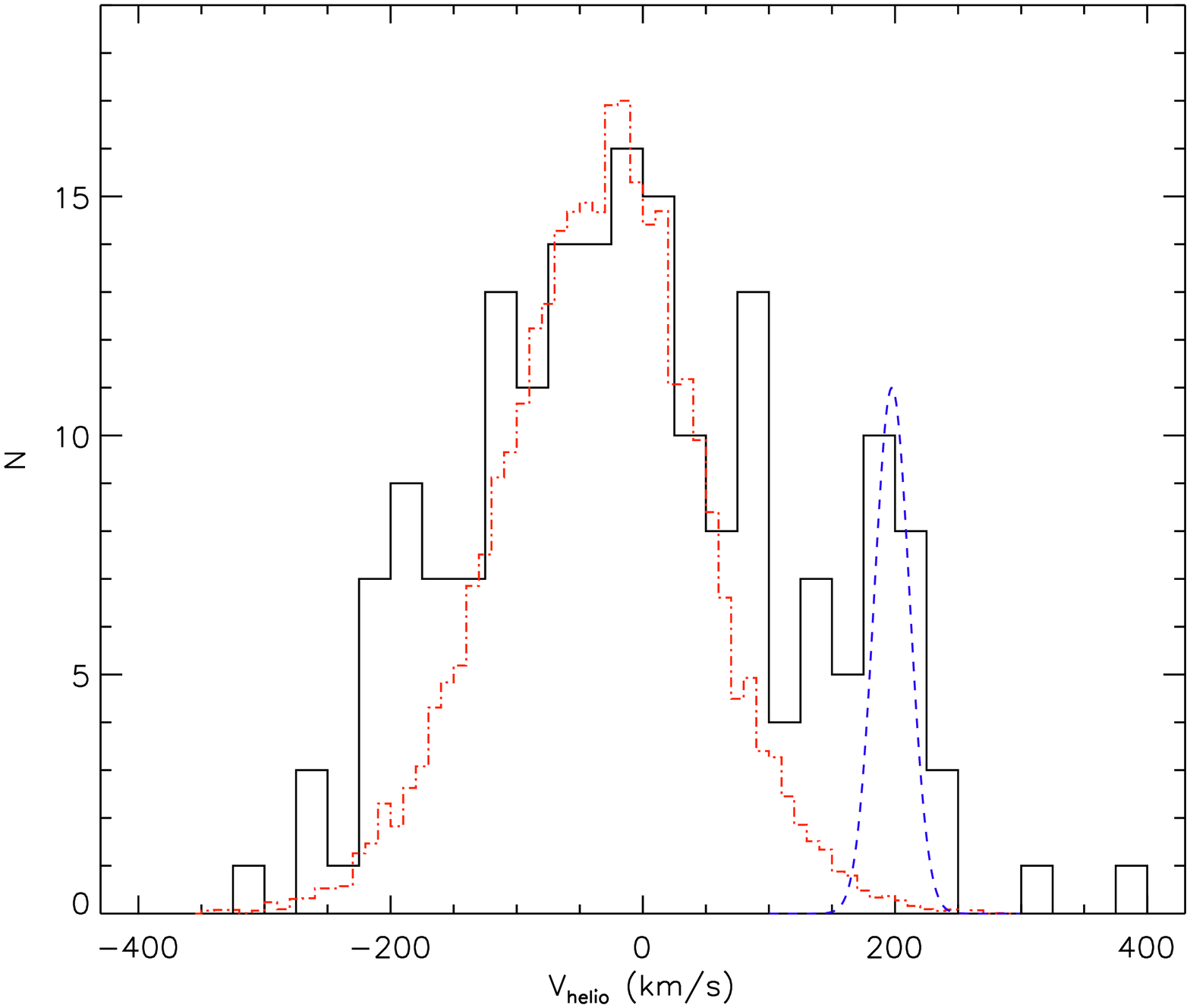}
\caption{Heliocentric radial velocities (solid histogram) for all
stars in the MMT+Hectospec configuration with reliable measurements.
For comparison we show scaled output from the Besan\c{c}on model as the
red dot-dashed histogram.  The model data consist of stars within 0.2 $<
g-r <$ 0.6, and 18.6 $< g <$ 22.0, which are predominantly Milky Way
halo turnoff stars in the same region of the CMD as our targets.
Relative to the model, there are two peaks of ``excess'' counts: one
at roughly -180 km s$^{-1}$, and a second at V$\sim$200 km s$^{-1}$.
We identify this latter peak as associated with BooIII.  The blue
dashed curve is a Gaussian with mean velocity $V_{\sun}=197.5$ km
s$^{-1}$ and a width of $\sigma_{\rm o}=14.0$ km s$^{-1}$, matching
our final derived values for BooIII.}
\end{figure}

Cross-correlation of BHB spectra in our sample against the two
available G-dwarf standard spectra results in RVs with higher
uncertainties than we derive for the later-type stars in our
sample. This is due to the mismatch in spectral type and spectral line
profiles between the BHB candidates and the RV standard stars.
Because of this, the scatter in our BHB RVs is likely artificially
high, potentially leading to higher field-star contamination in our
3-$\sigma$ velocity selection.  In order to reach a cleaner sample of
BooIII members, we further exclude stars that are more than 0.25
magnitudes from the isochrone shown in Figure 1.  This leaves our
final sample of 20 BooIII candidate members (14 if the BHB stars are
excluded), which is represented in Figures 1, 3, and 4 with large filled
symbols and asterisks.

\subsection{Metallicity}

Stellar atmospheric parameters for all observed stars other than the
BHB candidates are determined using low-resolution Lick spectral
indices (as defined by \citealt{Worthey1994}).  Measured indices for
all input spectra are placed onto the standard Lick index system of
\citet{Schiavon2007} using offsets derived from standard spectra
observed with the same telescope/instrument setup.  We derive a
suitable estimate of the effective temperature, $T_{\rm eff}$, of each
star based on its ($g - r$) photometric color.  Relatively accurate
dwarf/giant separation is then performed using the gravity-sensitive
Mg b feature at $\sim5170$ \AA\ , with stars sorting well by gravity
in the Mg b-H$\beta$ index plane (H$\beta$ is primarily sensitive to
$T_{\rm eff}$).  The metallicity, [Fe/H], for each star is determined
using two methods; from the \citet{Schiavon2007} spectral library we
have derived separate polynomial surface fits of (1) [Fe/H] to the
Lick iron indices and stellar color, and (2) [Fe/H] as a function of
the iron indices and H$\beta$ (similar to the \citealt{Friel1987}
interpolation between lines of constant metallicity, or the
\citealt{Gorgas1993} fitting functions).  Thus, moderately accurate
estimates of [Fe/H] derive from the combination of all six Lick iron
indices falling within the observed spectral bandpass as well as the
stellar color and H$\beta$ index.

Measured metallicities as a function of $V_{\rm helio}$ are shown in
Figure 3.  We estimate typical uncertainties in [Fe/H] of
$\sim$0.3-0.5 dex, depending on the magnitude of the star.  According
to \citet{Grillmair2009}, the matched-filter signal of BooIII is
strongest for the most metal-poor filter applied (M15: [Fe/H]=-2.26;
\citealt{Harris1996}).  Furthermore, the M15 ridgeline provides the
best fit to the CMD locus of BooIII (\citeauthor*{Grillmair2009},
Figure 12).  We thus expect the majority of BooIII members to be
metal-poor ([Fe/H] $\lesssim$ -2.0).  Examining the stars within the
3-$\sigma$ RV selection in Figure 3, we note that a number of
candidates are indeed rather metal-poor.  Furthermore, both of the
stars with S/N $>$ 20 spectra that were excluded from the final sample
as likely foreground contaminants (i.e. open triangles in Figures 1,
3, and 4) have measured [Fe/H] $>$ -1.8.

From the individual stellar metallicities of our final sample
(excluding the BHB stars) we derive a maximum likelihood estimation of
[Fe/H] $\approx$ -2.1 $\pm$0.2 for BooIII.  Since these individual
values are derived from low S/N spectra, we shift the BooIII
candidate spectra to rest wavelength and stack them to produce a
higher S/N spectrum.  Doing so, we derive [Fe/H] $\approx$ -2.0,
which is consistent with all other measurements for BooIII.  At the
integrated $V$-band absolute magnitude of BooIII, $M_{\rm
V}=-5.8\pm0.5$ \citep{Correnti2009}, our measured metallicity is
typical for a dwarf galaxy (see Figure 11 of \citealt{Simon_Geha2007}),
and more metal-poor than all Galactic globular clusters at similar
luminosity.  As is obvious from Figure 3, there is a fairly large
metallicity spread in the BooIII candidate stars ($\sigma_{\rm [Fe/H]}
\approx$ 0.6).  This may be due to foreground contamination of our
sample, though as seen in Figure 2, very few Galactic stars are expected
at these velocities.  This spread in [Fe/H] is comparable to (though
slightly higher than) those found by \citet{Simon_Geha2007} for a
number of other ultrafaint dwarfs (see their Table 4), and may
reflect the presence of multiple generations of star formation among
the BooIII stellar population.

\begin{figure*}
  \begin{center}
    \leavevmode
    \epsfxsize=16cm\epsfbox{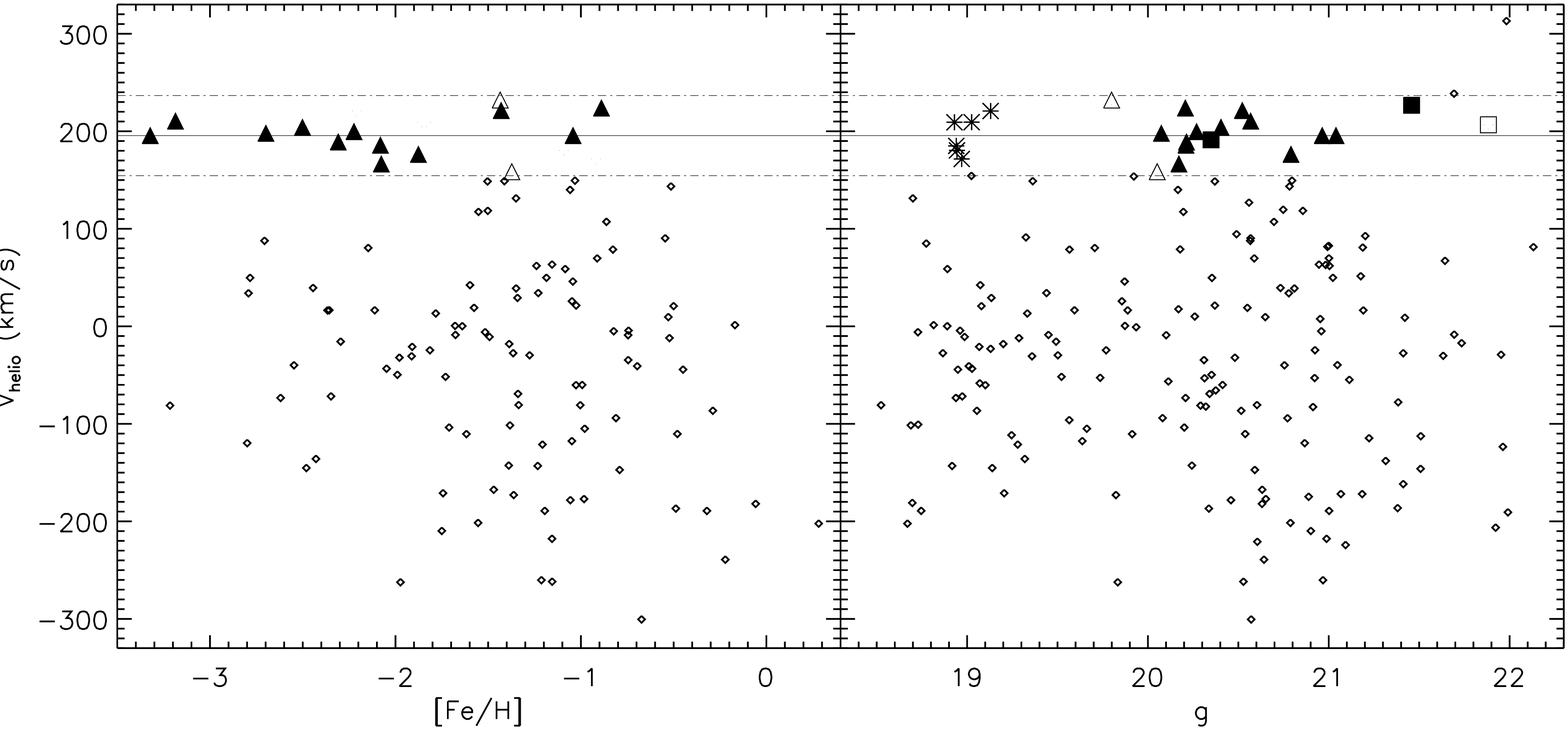}
    \caption{{\it Left panel:} heliocentric radial velocities as a
function of metallicity ([Fe/H]) for all stars in our observed sample
having spectra with S/N $>$ 20.  {\it Right panel:} RVs vs. SDSS $g$
magnitude for all stars with spectra.  Symbols for both panels are as
in Figure 1.  As in previous plots, open symbols are those stars that
were excluded from the final sample of BooIII candidates.  The
horizontal lines in both panels represent the mean (solid line) and
3 $\sigma$ (dot-dashed line) RV selection criteria based on the BHB
stars.}
  \end{center}
\end{figure*}


\subsection{Velocity Dispersion and Mass Estimate}

With our final sample of BooIII candidates, we use the maximum
likelihood method (\citealt{Pryor1993}; \citealt{Hargreaves1994};
\citealt{Kleyna2002}) to derive the systemic velocity and velocity
dispersion of Bo\"{o}tes III.  For the full sample (20 stars,
including the 6 BHB stars), we find $V_{\sun}=197.5\pm3.8$ km s$^{-1}$
and a velocity dispersion of $\sigma_{\rm o}=14.0\pm3.2$ km s$^{-1}$.
Removing the BHB stars (leaving 14 total), the result is virtually
unchanged: $V_{\sun}=198.2\pm4.5$ km s$^{-1}$, and $\sigma_{\rm
o}=14.1\pm3.7$ km s$^{-1}$ -- we thus consider only the full sample
henceforth.  At the position and distance of BooIII, this $V_{\sun}$
converts to an unusually high Galactic standard of rest velocity of
$V_{\rm GSR}=238.8\pm3.8$ km s$^{-1}$ which implies that BooIII is on
a very radial (and thus potentially destructive) orbit.  This idea of
a high-latitude object on a rather eccentric orbit agrees with the
suggestion by \citeauthor*{Grillmair2009} that BooIII may be extended
along the line of sight (i.e., along its orbit).

Our measured dispersion of 14 km s$^{-1}$ for BooIII is the highest
measured line-of-sight velocity dispersion of any dSph, even higher
than the central dispersion of Sagittarius, the most massive of the
Galactic dSph galaxies.  It has been shown in previous studies (e.g.,
\citealt{PP95}; \citealt{OLA95}; \citealt{Kroupa1997};
\citealt{Munoz2008}) that even if a satellite is experiencing tidal
disruption, its $\sigma_{\rm o}$ remains unaffected by tides {\it
unless} the system is at or near the point of complete destruction (as
we believe is the case for BooIII). In this context, it is possible
that even in the case of Sagittarius the central dispersion reflects
the equilibrium value.  The same studies, in particular
\citet{Munoz2008}, show that when a satellite is on the brink of being
completely destroyed by the action of tides, its central velocity
dispersion inflates in some cases by more than a factor of 2.  If
BooIII was originally a system similar to the classical dSphs, with a
half-light radius of $\sim300$ pc and a velocity dispersion of
$\sim6-7$ km s$^{-1}$, then its current measured value of 14 km
s$^{-1}$ is consistent with a scenario wherein BooIII is near complete
destruction.

In order to estimate the dynamical mass of BooIII we use a method
recently developed by \citet{2009AAS...21431105W} which relies solely
on the central velocity dispersion and the half-light radius of the
system, and assumes neither that the density distribution of the
satellite follows a King-like profile nor that mass follows light.  We
note that this does assume we are observing a bound system in
dynamical equilibrium, which we argue is {\it not} the case for
Bo\"{o}tes III.  For illustrative purposes, however, we carry out the
exercise.  The expression derived by Wolf et al. depends only on
direct observables, and is independent of the velocity anisotropy:
$M_{1/2} = \frac{3 \sigma^{2}_{\rm LOS} r_{1/2}}{G} \simeq \frac{4
\sigma^{2}_{\rm LOS} R_{\rm eff}}{G}$.  Here, $\sigma_{\rm LOS}$ is
the line-of-sight velocity dispersion, $r_{1/2}$ is the
three-dimensional deprojected half-light radius, and $R_{\rm eff}$ is
the two-dimensional projected half-light radius (using their derived
approximation $R_{\rm eff} \simeq (3/4) r_{1/2}$).  Substituting for
the constants, this can be expressed as M$ (< r_{1/2}) \simeq 9.3
\times 10^4 \frac{R_{\rm eff}}{100 \rm pc} (\frac{\sigma_{\rm
LOS}}{\rm km/s})^2$ M$_{\sun}$.  Taking our measured velocity
dispersion $\sigma_{\rm o}=14.0$ km s$^{-1}$, we find M$ (< r_{1/2})
\simeq 1.82 \times 10^7 \frac{R_{\rm eff}}{100 \rm pc}$ M$_{\sun}$,
where we've left $R_{\rm eff}$ explicit to allow for the as-yet
unmeasured surface brightness profile of BooIII.

This derived mass is rather typical for a dSph system --
\citet{Strigari2008} showed that Local Group dwarf galaxies all share
a common mass scale of the order of $\sim10^{7} M_{\sun}$ measured
within the inner $300$ pc of these satellites (as had earlier been
suggested by \citealt{Mateo1998}). Note that the above mass estimate
for BooIII would be three times higher if we take $R_{\rm eff} = $300 pc,
although the true $R_{\rm eff}$ is likely to be smaller
(\citeauthor*{Grillmair2009}).  Adopting $M_{\rm V}=-5.8$ gives a
total luminosity of $L = 1.8\times10^{4}$ L$_{\sun}$.  Using this
value along with our mass estimate, we derive a total mass-to-light
ratio of (M/L)$_{\rm V}\sim1000 \frac{R_{\rm eff}}{100 \rm pc}$
(M/L)$_{V,\sun}$.  A mass-to-light ratio of $\sim$1000 is similar to
those measured for the ultrafaint dSphs of similar $M_{\rm V}$
(e.g. \citealt{Simon_Geha2007}, Figure 15).

\section{Discussion}

Despite the seemingly ``normal" mass derived for BooIII we argue that
a number of pieces of evidence support the idea that BooIII is not in
dynamical equilibrium and it is likely in a transitional state between
a bound system and a completely unbound stellar stream -- similar to
the objects described by \citet{Kroupa1997} but with a high dark
matter content.  As we have already suggested, the fact that BooIII has
the highest measured line-of-sight velocity dispersion of any dSph is
consistent with the observation (from, e.g., \citealt{Munoz2008}) that
a satellite in the throes of complete tidal disruption can have its
velocity dispersion inflated by more than a factor of 2.  Tidal
disruption seems quite plausible, since the large measured
Galactocentric velocity for BooIII places it in a highly elongated,
and thus potentially destructive, orbit.

\begin{figure}
\plotone{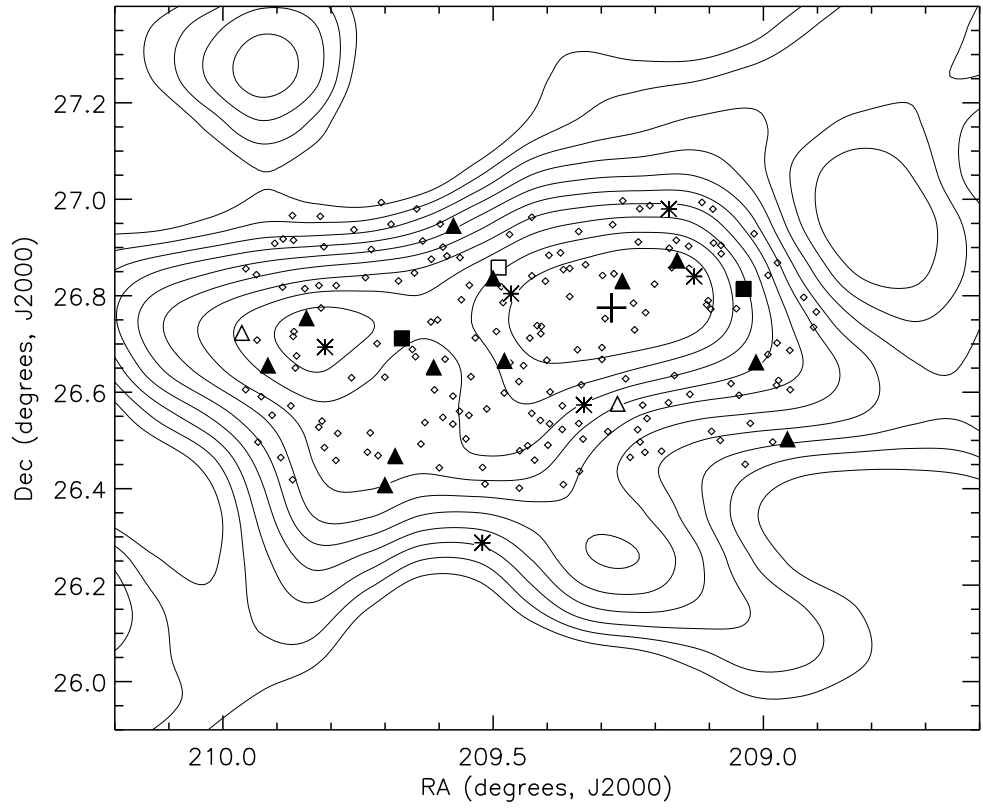}
\caption{Spatial distribution of all observed spectroscopic targets.
Symbols as in Figure 3.  Contours represent filtered star counts, and
the large plus sign denotes the BooIII center, both from the
\citeauthor*{Grillmair2009} study.}
\end{figure}

Morphological evidence further supports the suggestion that Bo\"{o}tes
III is a dissolving dwarf galaxy in the final stages of tidal
destruction.  The density distribution derived by
\citeauthor*{Grillmair2009} (and confirmed by \citealt{Correnti2009})
is quite unusual.  Most dSphs, with perhaps the exception of a few of
the new ultrafaint ones which are suspected to be undergoing tidal
stripping, have density profiles that are well described by King-like
or Plummer-like profiles.  In contrast, Figure 11 of
\citeauthor*{Grillmair2009} (and discussion by \citealt{Correnti2009})
shows that BooIII is better described by a power law of index -1. This
observational result is also consistent with the idea of a satellite
on the brink of destruction. Several studies (e.g.,
\citealt{Johnston_etal1999}; \citealt{Munoz2008};
\citealt{Penarrubia_etal2008}) have shown that a satellite which
started with a King-like density distribution develops a power
law-like distribution in the outer parts of its luminous component as
it undergoes tidal stripping (due to escaping stars at large radii).
Eventually, and as the satellite becomes nearly unbound, the King-like
core also turns into a power-law like distribution.  Furthermore,
Grillmair (2009; Fig. 10) showed that BooIII has irregular
stellar surface density contours and is extended along the east-west
direction.  In Figure 4, we show the spatial distribution of our
spectroscopic targets overlaid on contours of the
\citeauthor*{Grillmair2009} filtered star counts.  Though our single
MMT+Hectospec setup covers only a fraction of the rather large BooIII
stellar overdensity, the identified RV members seem to
follow the BooIII contours.  The Styx stellar
stream also passes through the same region of the sky where BooIII is
located, and at nearly the same distance
(\citeauthor*{Grillmair2009}).  BooIII may be associated with this
stream, and indeed may even be its progenitor.  However, further
kinematical information along the Styx stream will be necessary to
confirm the association.

We thank the anonymous referee for helpful suggestions, and gratefully
acknowledge support by NSF grant AST-0807945 and NASA/JPL contract
1228235.

\end{document}